\documentclass [10pt]{article}
\usepackage{times}
\usepackage[dvips]{graphicx}
\newcommand{\dd}{\mbox{$\nabla^2$}}
\newcommand{\strayer}{\it Strayer University \\
One Penn Center West, Pittsburgh, PA 15276}
\newcommand{\auth}{ M. Mihailescu
\footnote{Email address: mhl@strayer.edu }}
\begin{document}
\hsize=126mm
\vsize=180mm
\parindent=5mm
\def\lhead{ M. Mihailescu } %% for author's initials and surname
\def\rhead{Higher derivative free energy terms and interfacial curvatures} %% for abbreviated title

\begin{center}
{\large {\bf Higher derivative free energy terms and interfacial curvatures}}

\vspace{30pt}
\auth

%\vspace{10pt}
{\strayer}
     
\vspace{40pt}
         
\begin{abstract}
High derivative terms do not play a major role in field theories because of the associated complexity and inherent difficulty in connecting these terms to physically measurable quantities. A role for higher derivative terms is analyzed for the case of field theories used to describe phase separated systems. In these theories, higher derivative terms are directly connected to an interfacial free energy which contains the mean and the Gaussian curvature and are shown to determine explicitly the shape of the interface.          
\end{abstract}

\end{center}

\newpage

\section{Introduction} 
Cooperative behavior of lipid molecules in biological membranes is believed to play a central role in processes happening at cellular level. More generally, cooperative processes appear in systems exhibiting hydrophobic-hydrophilic interaction even beyond those appearing in biology \cite{Chandler}. Cooperative processes are captured and studied with difficulty by numerical simulations at molecular level, mainly because those methods have a low probability of capturing events involving multiple correlated molecular steps. Moreover, simulations at molecular level are usually plagued by an extensive range of parameters and although successful in predicting short-time molecular dynamics, they make less of an impact in a realistic statistical system that covers large spatial and temporal scales. Other computational approaches are bound to improve the relatively straightforward and parameter-extensive all-atom simulations. These methods made their way towards molecular simulations \cite{Muller} and they appear under the terminology of coarse-grained numerical methods. Coarse-grained methods are based on a considerably reduced parameter space and provide a significantly increased speed. Therefore, coarse-grained methods cover the long-time dynamics familiar to many cell-biology cooperative processes (nanoseconds to seconds) with the additional advantage of allowing an intrinsic inclusion of statistical averaging. In spite of the advantages, coarse-grained methods present challenges of their own. For example, in the framework of these methods, making contact with realistic systems and identifying those few most relevant parameters is usually a difficult task.   

Particular chemical systems where cooperative behavior is responsible for a geometrical rich phase structure are polymeric blends.  For this case, the cooperative behavior is due to the long-range interaction stemming from "stretching" abilities of polymer chains. A coarse-grain computational and theoretical model where molecular details and coordinates are replaced by a field theory of concentrations is already applied and used for a while \cite{helfand}, \cite{Matsen}. Self-consistent field theory methods which are based on a mean field approach provide a successful representation of the experimental data. These methods were recently applied with success to biological membranes and other complex cellular structures, such as vesicle, see for a review \cite{MembraneReview}. One of the major drawbacks of the "self-consistent" methodology is the lack of transparency. In particular, a direct relation to physical measurable parameters, such as interfacial tension and geometrical invariants, such as mean and Gaussian curvature of interfaces is difficult to write explicitly. Moreover, the method faces difficulties when extended to dynamical models \cite{fraaije}, difficulties similar in magnitude to those appearing in molecular models.  

A more explicit field theory model can be obtained for a system of homopolymers, using a formal perturbation expansion of the corresponding self-consistent field theory. The result of the expansion is a regular free energy expression that contains higher derivative free energy terms \cite{freed}. This type of terms come regularly in effective field theories from integration of higher energy modes (or alternatively of quickly relaxing modes). Although these terms appear in other field theories used to describe strongly correlated systems \cite{nussinov}, they are not regularly treated mainly because the complexity and lack of transparency associated with them. It is the goal of this work to clarify the connection between higher derivative free energy terms for a phase separated system and the coefficients that govern the dynamics of interface geometry. 
\newpage 

A common phenomenological form for the interfacial action was proposed by Helfrich \cite{helfrich}:
\begin{equation}
\label{HelfrichS}
F_S=\int d\Sigma(\gamma + c_H (H-H_0)^2 + c_G K )
\end{equation}
where $\gamma$ is the interfacial tension, $c_H$ is the free energy coefficient associated with mean curvature $H$, $H_0$ is the intrinsic mean curvature and $c_G$ is the free energy coefficient associated with Gaussian curvature $K$. Such an interfacial free energy expression is useful both from the experimental and theoretical point of view, but the above free energy can be used to describe a static interface and does not allow a continuous representation for common experimental processes such as joining and splitting of interfaces. In this study, starting from a generic free energy that contains additional higher derivative terms, the direct contact with the free energy (\ref{HelfrichS}) is made. In addition, it is shown that the interfacial free energy naturally features contributions from the interfacial width, contributions that can be directly correlated with the disappearance or formation of interfaces. Another advantage is that from the point of view of the number of degrees of freedom, a complete matching between bulk and interfacial degrees of freedom is achieved. Matching of degrees of freedom is reached through the usual scheme of dimensional transmutation, where a coordinate, the one that is normal on the interface, is transformed in an additional scalar field that lives on the interface, the interfacial width. The picture is not yet complete, but the interfacial width presents itself as the required intermediary field that provides a continuous description to the nontrivial topological change associated with joining and splitting. The interfacial free energy thus obtained reveals also nontrivial interactions between the interfacial width, and the geometrical properties of the interface, such as mean and Gaussian curvature. Such interaction terms come from higher derivative terms in free energy and open the possibility of modulating the interface itself, in particular the interfacial width, by varying the curvature of the interface.     

The plan of the paper is as follow. In section 2, the method of derivation for the interfacial free energy from a generic free energy action is presented. In section 3, the method is applied to the case of Landau-Ginzburg $\phi^4$ potential and all interfacial coefficients are calculated. The general case is presented in section 4 and section 5 contains the calculation of interfacial parameters for the realistic case of binary incompressible blend of homopolymers with different molecular weights.

\section{From higher derivatives to interfacial properties}

Higher derivative terms appear naturally in effective free energy actions for statistical systems and their appearance stems from an integration of degrees of freedom with relaxation time scales shorter than the scale time of interest (namely, those degree of freedom which are inaccessible to the particular experimental setup). Alternatively, these terms can also appear in quantum effective actions, but here, their appearance is a sign of the underlying intrinsic quantum dynamics of fields. The problem becomes more involved when the free energy admits soliton-type solutions, solutions that in a traditional physics setup excite both low and high energy modes and it is not always clear how to separate (and integrate) the energy modes. 

In this work, it is assumed that following an integration of inaccessible modes, an effective free energy with higher derivative terms is already available. This effective free energy still possesses an effective potential that admits domain - wall solutions (soliton-like solutions). Within this setup, a free energy admitting domain-wall solutions and having also higher derivative terms, the impact of higher derivatives terms on the physical characteristics of the domain-wall solution is analyzed. Although such a free energy may look esoteric, it is shown in section 5 that such expressions can be obtained in realistic systems such as an incompressible mixture of homopolymers. 

The generic free energy expression considered in this work is: 
\begin{eqnarray}
F[\phi] & = & F_1[\phi]+ F_2[\phi] \label{action} \\
F_1[\phi] & = & \int d^3 r \Big( V(\phi)+ M(\phi)( \nabla \phi)^2 \Big) \label{action1}\\
F_2[\phi] & = & \int d^3 r \Big( N(\phi)( \nabla \phi)^4 + P(\phi)( \nabla \phi)^2 ( \dd \phi)+ Q(\phi) \partial_i \phi \partial_j \phi \partial^{ij} \phi + \nonumber \\ & & \qquad \quad + R(\phi)(\dd \phi)^2 + S(\phi) \partial_{ij} \phi  \partial^{ij} \phi \Big) \label{action2}
\end{eqnarray}
where $F_1$ in equation (\ref{action1}), is the low energy part of free energy and $F_2$, in  equation (\ref{action2}) is the correction part of the free energy coming from integration of degrees of freedom and contains higher derivatives. $F_2$ is considered to formally be a perturbation effect added to $F_1$. $V(\phi)$ is a positive potential that has two minima $\Psi_{-,+}$ corresponding to the two phase components of the system. The value of the potential in those phases, namely at minima points $\Psi_{-,+}$, is considered to be zero. Under this assumption, the free energy density in the bulk of each phase, namely in the regions where the field is constant and equal with either $\Psi_{-}$ or $\Psi_{+}$, is $0$. Thus, no contributions to the free energy come from bulk regions and in fact, all free energy contributions come from the spatial region where the values of the field varies between the two coexistence values $\Psi_{-,+}$. These suggest that a complete reduction of free energy to such a region can be made giving rise to a free energy defined on a one dimensional lower space, namely defined on the interface. In order, to employ such a reduction, an Ansatz using the domain-wall solution of the equation of motion for the first two terms of the free energy, $F_1$ is used. This equation of motion is:
\begin{equation}
\label{FirstEq}
-2 M(\phi) {d^2 \over {d x^2}} \phi - M'(\phi) ({d \over dx} \phi)^2 +V'(\phi)=0
\end{equation}
where the $'$ denotes the first order derivative in $\phi$. Following the usual procedure for domain-walls, this equation can be reduced to a first order differential equation: 
\begin{equation}
\label{DomEq}
{d \over dx} \phi=\left({V(\phi) \over M(\phi)}\right)^{1/2}
\end{equation}
The solution to this equation that describes a domain wall with an interface at $x=0$ can in principle be determined and denoted as $\Psi(x)$. This solution goes from $\Psi_{-}$ at $x=-\infty$ to $\Psi_{+}$ at $x=+\infty$. The maximum of the first derivative of $\Psi(x)$ is taken to be at $x=0$ and this point is considered to be the interface location. The following field redefinition is proposed:
\begin{equation}
\label{redef}
\phi(r)=\Psi(f(\bf{r}))
\end{equation}
where $f(\bf{r})$ is a field with dimension of length. The field $f(\bf{r})$ represents physically the geometric profile of the interface, and it is used to describe the location of the interface in three dimensions through the equation $f(\bf{r})=0$. The free energy is rewritten in terms of the geometrical field $f(\bf{r})$ used in (\ref{redef}) using the relations below between derivatives of field $\phi$ and those of field $f$:
\begin{eqnarray}
\partial_\alpha \phi & = & \Psi'(f(r))\partial_\alpha f(r)\\
\partial_{\alpha \beta} \phi & = & \Psi'(f(r))\partial_{\alpha \beta} f(r) + \Psi''(f(r))\partial_\alpha f(r) \partial_\beta f(r)
\end{eqnarray} 
These equations are substituted in equation (\ref{action}). A simplification in calculations is brought by expressions that relate the derivatives of $\Psi'(f)$ and $\Psi''(f)$ in terms of $\Psi(f)$ obtained using the equation (\ref{DomEq}) and equations derived by its differentiation:
\begin{eqnarray}
\Psi'  & = & \left(V(\Psi) \over M(\Psi) \right)^{1 \over 2} \label{deriv1} \\
\Psi''  & = &  { 1 \over 2}  \left( { {V'(\Psi) M (\Psi) - V (\Psi) M'
(\Psi) } \over M(\Psi)^2} \right) \label{deriv2}
\end{eqnarray}
where $\Psi$ depends on $f$ and represents the domain-wall profile. In terms of the field $f(\bf{r})$ and its derivatives the free energy can be finally written as: 
\begin{eqnarray}
\label{actionF}
F[f] &=&  F_1[f] + F_2[f] \label{actionF} \\
F_1[f] &=& \hspace{-10pt}\int d^3 r M(\Psi) \Psi'^2 \Big(1 + (\nabla f)^2\Big) \nonumber\\
F_2[f] &=& \hspace{-10pt} \int d^3 r \bigg( \Big( N(\Psi) \Psi'^4 + (P(\Psi) + Q (\Psi) ) \Psi'^2 \Psi''+ ( R(\Psi) + S (\Psi) ) \Psi''^2 \Big) (\nabla f)^4 + \nonumber \\
&& \hspace{-45pt}+ \Big(P(\Psi) \Psi'^3 +  2 R(\Psi) \Psi' \Psi'' \Big) (\nabla f)^2 \dd f + \Big( Q(\Psi) \Psi'^3 + 2 S(\Psi) \Psi' \Psi'' \Big) \partial_\alpha f \partial_\beta f \partial^{\alpha\beta} f + \nonumber \\
&& \hspace{-45pt} + R(\Psi) \Psi'^2 (\dd f)^2 + S(\Psi) \Psi'^2 \partial_{\alpha\beta} f \partial^{\alpha\beta} f \bigg)  \nonumber
\end{eqnarray}
It is worth mentioning that in the equation above, the dependence on $f$ comes both explicitly, as seen above, and implicitly through the profile function $\Psi(f)$. The free energy in terms of the field $f$ are similar in terms of derivative term representation with the initial free energy. Further simplifications are obtained by choosing a particular Ansatz for $f$. The choice in $f$ removes all geometrical ambiguities related with describing the surface through a functional equation of type $f(\bf{r}) = 0$ and allows the introduction of interfacial physical parameters in the free energy. A local system of coordinates is used to simplify the calculations of interfacial free energy terms. The derivation is done for a particular interfacial point and a particular system of coordinates. The expected geometrical invariance of interfacial free energy is used then to generalize the local free energy to a geometrically invariant form, valid at any interfacial point and in any coordinate system. The analysis is carried in three dimensions, but it can be observed that most of the features can be generalized directly to any dimension. By translation and rotation transformations, the given interfacial point is considered to be $(0,0,0)$ and the Ansatz for $f(x,y,z)$ around this point is: 
\begin{eqnarray}
f(x,y,z)&=& \lambda(y,z) \Big(x - \sigma(y,z)\Big) \label{local} \\
\sigma(0,0)&=&0  \nonumber \\
\partial_{y,z} \sigma(0,0)&=& 0 \nonumber \\ 
\lambda(0,0) \geq 0 \nonumber 
\end{eqnarray}
where $x=\sigma(y,z)$ is the location of the interface in the new parameterization and $\lambda(y,z)$ is a scalar field on the interface that can take only positive values and is directly proportional with the inverse of the interfacial width. Such a choice for $f(x,y,z)$ although apparently limiting, in fact, does not make any restriction on the interface. Therefore, the choice in (\ref{local}) represents the most general interface in a particular local coordinate system valid in the neighborhood of the point $(0,0,0)$. 
The choice can be justified by trying to fix the large freedom in choosing the curve equation as a function of regular spatial coordinates. For example, the interfacial function $f$ is in first approximation linearly proportional to the normal coordinate $x$ at the interfacial point $(0,0)$. The final form for $f$ in equation (\ref{local}) is obtained through translation and rotation of the local coordinate system such that $x$ axis is normal on the surface $x=\sigma(y,z)$, whereas $y$, $z$ axes are parallel to the tangent plane of interface at interfacial point $(0,0)$. The free energy density is computed in the neighborhood and the limit $(y,z)\rightarrow(0,0)$ is used freely in the computation. Overall, the procedure outlined above amounts to reduction of the free energy density along the direction normal to the interface using the known expression of the one dimensional profile. As it is shown in the appendix 1, at point $(0,0)$ only, the metric in parameterization (\ref{local}), is the trivial two dimensional metric. The second fundamental form is given in terms of second order derivatives of $\sigma$, whereas the mean curvature is given by $\dd \sigma$. 
In order to fully reduce the three dimensional free energy to a two dimensional interfacial free energy, it is necessary to absorb the normal direction $x$. This is accomplished by integration of free energy density at each point of the interface, on normal coordinate to the interface, $x$. The coordinates $y$, $z$ describe the interface and the assumption is made that after integration a local interfacial density is obtained. A straightforward integration over the normal coordinate, $x$ is done by a change of variables in a neighborhood of the point $(0, 0) $from $x$ to $\it{f}$: $x={\it{f} \over \lambda(y,z)} + \sigma(y,z)$. At the interfacial point, $\sigma(0,0)=0$ and $x = {\it{f} \over \lambda(0,0)}$.  This change of coordinates leads to the following expression for integration ($(y,z) \rightarrow (0, 0)$): 
\begin{equation}
\label{var}
\int dx dy dz ...\longrightarrow \int dy dz \int {d\it{f} \over \lambda(y,z)} ...
\end{equation}
Integration over $\int df$ is made first at the point $(y,z)=(0,0)$ and it is followed by integration over the interface coordinates $y$ and $z$. As it was mentioned above, the integrand expression obtained at $(0,0)$ in the chosen local coordinate system is generalized using the expected geometrical invariance of the final expression of free energy density. Thus, the trivial metric and measure at point $(0,0)$ becomes the full metric on the interface, $\hat{g}$ with $g_{\alpha \beta}$ ( $\alpha$ and $\beta$ are interfacial coordinates, e.g. $y$ and $z$) and measure $ d\Sigma = \sqrt{\det \hat{g}} dy dz$. 
\newpage
For completeness, the expressions obtained for each term in equation (\ref{action}) are given below:
\begin{eqnarray}
F &=& \int d\Sigma {1 \over \lambda} \int_{-\infty}^{+\infty} df \bigg( M(\Psi) \Psi'^2 \Big( 1 + \lambda^2 + f^2 {(\nabla \lambda)^2 \over \lambda^2}\Big) + \nonumber \\
& & \hspace{-40pt} + \Big( N(\Psi) \Psi'^4 + \Big( P(\Psi) +Q(\Psi) \Big) \Psi'^2 \Psi'' + \Big( R(\Psi)+S(\Psi) \Big) \Psi''^2 \Big) \Big( \lambda^2+ f^2 {(\nabla \lambda)^2 \over \lambda^2} \Big)^2 + \nonumber \\
& & \hspace{-40pt} + \Big( P(\Psi) \Psi'^3 + 2 R(\Psi)\Psi' \Psi'' \Big) \Big(\lambda^2+ f^2 {(\nabla \lambda)^2 \over \lambda^2} \Big) \Big(f {\dd \lambda \over \lambda} - 2 \lambda H \Big) + \nonumber \\
& & \hspace{-40pt} + \Big( Q(\Psi) \Psi'^3 + 2 S(\Psi)\Psi' \Psi'' \Big) \Big( f (\nabla \lambda)^2 + f^3  {{\partial_{\alpha}\lambda \partial_{\beta} \lambda \partial^{\alpha\beta}\lambda} \over \lambda^3} - f^2 {{\partial_{\alpha} \lambda \partial_{\beta} \lambda} \over \lambda} \Pi^{\alpha\beta} \Big) + \nonumber \\
& & \hspace{-40pt} + R(\Psi) \Psi'^2 \Big(f {\dd \lambda \over \lambda} - 2 \lambda H\Big)^2 + \nonumber \\
& & \hspace{-40pt} + S(\Psi) \Psi'^2 \Big( f^2 {{\partial_{\alpha \beta} \lambda \partial^{\alpha \beta} \lambda} \over \lambda^2} + 2 (\nabla \lambda)^2 - 2 f \partial _{\alpha \beta} \lambda \Pi^{\alpha \beta} + \lambda^2  \Pi_{\alpha \beta} \Pi^{\alpha \beta} \Big) \bigg) \label{FreeS}
\end{eqnarray}
where $\Psi$ depending on $f$ is the domain-wall solution in the absence of higher derivative corrections, $\Psi'$ is its first derivative and $\Psi''$ is its second order derivative (see equation (\ref{redef}), (\ref{deriv1}), (\ref{deriv2})), $\lambda$ depending only on interfacial coordinates, e.g. $y$ and $z$, is the inverse of interfacial depth at the interfacial point $(y,z)$. $\Pi^{\alpha \beta}$ is the second fundamental form of the surface as embedded in the three dimensional space and is related with mean curvature $H={1 \over 2} tr(Pi)$ and Gaussian curvature $K =\det(\Pi)$ through equations:
\begin{eqnarray}
H &=&  {1 \over 2} \Pi_{\alpha} ^{\alpha} \label{HK1}\\
4 H^2 - 2 K & =& \Pi_{\alpha \beta} \Pi^{\alpha \beta} \label{HK2} 
\end{eqnarray}
(see also appendix 1). In the above expression, all derivatives are understood in terms of geometrical derivatives in interfacial coordinates, namely $\partial_\alpha$. The dependence of $\lambda$ and interfacial geometry is completely featured and only the corresponding values for coefficients are still to be determined by integration over $f$. The dependence on $f$ in the integrand comes directly and indirectly through the functions $\Psi(f)$, $\Psi'(f)$ and $\Psi''(f)$. Solving for the exact contribution to interfacial terms from each term implies integrating over $f$ and that can be done in the case when the exact expression of interfacial $\Psi(f)$ is known, namely for simple forms of potential $V(\phi)$, see section 3. Moreover, calculating the coefficients can also be made in other particular cases, when the potential is more complex and an exact analytic expression for $\Psi(f)$ is not possible, see section 5. An interesting feature of equation (\ref{FreeS}) is that it pinpoints the contribution of each of the terms from (\ref{action}) to the interfacial free energy (\ref{HelfrichS}. The summary of these dependencies is shown below for each geometrical term appearing in (\ref{FreeS}) 
\begin{eqnarray}
&& M(\phi), \quad N(\phi), \quad P(\phi), \quad Q(\phi), \quad R(\phi), \quad S(\phi) \longrightarrow  \lambda  \nonumber \\
&& P(\phi), \quad R(\phi) \longrightarrow  H  \nonumber\\
&& R(\phi) \longrightarrow  H^2  \nonumber\\
&& Q(\phi), \quad S(\phi) \longrightarrow  \Pi_{\alpha \beta} \nonumber \\
&& S(\phi) \longrightarrow  \Pi_{\alpha \beta} \Pi^{\alpha \beta} \nonumber
\end{eqnarray}
where the first expression shows the contributions to terms containing only $\lambda$, whereas the other expressions show contributions to terms proportional with $H$, $H^2$ and so on. The terms multiplied by $M(\phi)$ and $N(\phi)$ do not bring curvature terms in the interfacial free energy. Terms proportional with $K$ come only from $S(\phi)$ term, with $H^2$ from $S(\phi)$ and $R(\phi)$ terms, whereas the intrinsic curvature $H_0$ is modulated by $P(\phi)$, $Q(\phi)$, $R(\phi)$ and $S(\phi)$.     
The expression (\ref{FreeS}) shows that in addition to the geometry of the interface, the free energy density also depends on a scalar field, the inverse of interfacial width $\lambda$. The geometric invariants that make their way in the free energy explicitly at this level of approximation are those enclosed in the first fundamental form (metric) and second fundamental form, namely $\Pi^{\alpha \beta}$, or indirectly in the better known invariants, mean curvature $H$ and Gaussian curvature $K$. Since the goal of this work is to make contact with the interfacial free energy described in equation (\ref{HelfrichS}), only the terms of order $\nabla^4$ are considered (\ref{action}). It is worth mentioning that the same method can be applied for even higher derivative terms, e.g. of order $\nabla^6$, resulting in the appearance of other geometrical invariants in the interfacial free energy.

\section{Simplified example - Symmetrical double-well potential}
The expression developed in the previous section is applied to a simplified case. In this case a Landau-Ginzburg type free energy with a symmetric double-well potential, $V(\phi)$, is considered and higher derivative terms are added. Furthermore, for this section coefficients in the free energy equation (\ref{action}) are constant functions. In this case, the terms that come multiplied by $Q$ and $S$ are in fact redundant, their contributions being absorbed, up to total derivatives, in $P$ and $R$ constants. Also for clarity, the one dimensional domain-wall solution considered as reference in equation (\ref{redef}) is taken as simple as possible:  
\begin{equation}
\label{redef1}
\Psi(x)=\tanh(x)
\end{equation}
This expression for the domain-wall solution is consistent with values of the coefficients in $M$ and $V(\phi)$ shown below. The free energy expression considered in this section is:
\begin{eqnarray}
V(\phi) & = & {1\over 2}(\phi^2-1)^2; \quad\quad M(\phi)  =  {1\over 2}; \quad\quad Q = 0; \quad\quad S = 0 \label{P4action} \\
F[\phi] & = &  \int d^3 r \Big( {1\over 2}(\phi^2-1)^2+ {1\over 2}( \nabla \phi)^2 + N (\nabla \phi)^4 + P( \nabla \phi)^2 ( \dd \phi)+ R(\dd \phi)^2  \Big) \nonumber
\end{eqnarray}
Equations (\ref{FreeS}) and the explicit form for domain-wall solution, equation (\ref{redef1}) allow the explicit integration along the normal direction over $f$. The full derivation is shown in the appendix 2 and only final results are given below. The first two terms in the free energy expression (\ref{P4action}) lead after field redefinition, geometric identifications and integration over $f$ to:
\begin{equation}
\label{NoCorr}
\int dy dz \Big( {2 \over 3} {{1+\lambda^2} \over \lambda} +{{\pi^2 -6} \over 18} {(\nabla \lambda)^2 \over \lambda^3}\Big)  
\end{equation}
The term multiplied by coefficient $N$ leads to:
\begin{equation}
\label{NT}
N \int dy dz \Big( {32 \over 35}  \lambda^3 +{{8 (6\pi^2 - 49)} \over 315} {(\nabla \lambda)^2 \over \lambda}+{{2 (3 \pi^4 -35 \pi^2 +60)}\over 225}{(\nabla \lambda)^4 \over \lambda^5}   \Big)  
\end{equation}
and the term multiplied by coefficient $P$ shows the following contribution:
\begin{equation}
\label{PT}
P \int dy dz \Big( - {32 \over 15}  \lambda^2  +{2 (30 - 4\pi^2) \over 45} {(\nabla \lambda)^2 \over \lambda^2}  \Big) H  
\end{equation}
Finally, for the term whose proportionality constant is R, after integration over $f$ and cancellation of total derivative terms, the following expression is obtained:
\begin{eqnarray}
R \int dy dz \Big({16 \over 15}  \lambda^3  +{8 \pi^2 \over 45} {(\nabla \lambda)^2 \over \lambda} + { {7 \pi^4 -360} \over 225} {(\nabla \lambda)^4 \over \lambda^5} + \nonumber \\ + { {6 - \pi^2} \over 3} {{(\nabla \lambda)^2 \dd \lambda} \over \lambda^4} +  { {\pi^2-6} \over 9} {(\dd \lambda)^2  \over \lambda^3} + {16 \over 3} \lambda H^2  \Big) \label{RT}  
\end{eqnarray}

After collecting all terms from equations (\ref{NoCorr}), (\ref{NT}), (\ref{PT}) and (\ref{RT})and generalizing the expression from the neighborhood of interfacial point $(0,0)$ to any point on the interface, the following expression is obtained:
\begin{eqnarray}
F_S[\lambda, \sigma] & = & \int d\Sigma \bigg( \Big( {2 \over 3} {{1+\lambda^2} \over \lambda} + ( N  {32 \over 35}  + R {16 \over 15} ) \lambda^3 \Big) + \label{FS} \nonumber\\
& & + {(\nabla \lambda)^2 \over \lambda} \Big({{\pi^2 -6} \over {18 \lambda^2}} + N {{8 (6\pi^2 - 49)} \over 315} + R {8 \pi^2 \over 45} \Big)+ \nonumber \\
&&+ {(\nabla \lambda)^4 \over \lambda^5} \Big( N {{2 (3 \pi^4 -35 \pi^2 +60)}\over 225} + R { {7 \pi^2 -360} \over 225} \Big)+ \nonumber \\
&& + R  { {6 - \pi^2} \over 3} {{(\nabla \lambda)^2 \dd \lambda} \over \lambda^4} + R { {\pi^2-6} \over 9} {(\dd \lambda)^2  \over \lambda^3}+ \nonumber \\
&&+  P \, H \Big(- {32 \over 15}  \lambda^2 +{2 (30 - 4\pi^2) \over 45} {(\nabla \lambda)^2 \over \lambda^2} \Big)+ R {16 \over 3} \, \lambda H^2  \bigg)   
\end{eqnarray}
  
The above explicit interfacial free energy expression shows the nontrivial character of the higher derivative terms for the case of a simplified potential. It shows a rational dependence on interfacial width, which can not be obtained through symmetry or other qualitative arguments. Furthermore, the free energy for this particular case does not depend on Gaussian curvature, but only on mean curvature. The independence of free energy on Gaussian curvature is consistent with the fact that $S$ is constant function and with the topological character of two dimensional Gaussian curvature. Gaussian curvature appears only when a more involved function is chosen as $S(\phi)$ in equation (\ref{action}). Lastly, the expression (\ref{FS}) shows explicitly the coupling between the geometric parameters such as mean curvature and interfacial width suggesting that it is possible to affect interfacial width, e.g. extend or contract, by manipulating the curvature of the interface. More detailed conclusions will be given in the next section and in section 5, in the case of homopolymers blend where entropic free energy contributions for internal degrees of freedom for an extended object are responsible for the higher derivative terms. 

\section{Interfacial free energy - General case}
To gain a better understanding for the interfacial free energy, a few assumptions are made. The expression (\ref{FreeS}) is simplified to a level that allows an analysis of the effects of higher derivative terms. The main assumption is that the interfacial width does not vary across the interface, meaning that $\nabla \lambda = 0$ and $ \dd \lambda =0$ and that results in the elimination of such terms from free energy expression (\ref{FreeS}). This assumption restricts the type of systems where the method can be applied. The assumption is also satisfied for a general system, but in cases when the interface has an intrinsic symmetry. For example, planar, spherical, cylindrical or toroidal interfaces in three dimensions have such symmetries. Under these conditions, the free energy expression becomes:
\begin{eqnarray}
F & = & \int d\Sigma {1 \over \lambda} \int_{-\infty}^{+\infty} df  \bigg( M(\Psi) \Psi'^2 \Big( 1 + \lambda^2 \Big) + \Big( N(\Psi) \Psi'^4 + \Big(P(\Psi) +Q(\Psi)\Big) \Psi'^2 \Psi'' + \nonumber\\
&& + \Big(R(\Psi)+S(\Psi)\Big) \Psi''^2 \Big) \lambda^4 -2 \Big( P(\Psi) \Psi'^3 +2 R(\Psi)\Psi' \Psi'' \Big) \lambda^3 H + \nonumber \\
&& + 4 R(\Psi) \Psi'^2 \lambda^2 H^2 + 4 S(\Psi) \Psi'^2 \lambda^2 \Big (H^2-{K \over 2} \Big) \bigg) \label{FreeSS}
\end{eqnarray}
where equations (\ref{HK1}), (\ref{HK2}) are used to provide transparency. The assumption taken at the beginning of the section leads to a free energy density that does not depend on $f$ explicitly, but through $\Psi(f)$. It is possible now to avoid solving the domain-wall equation (\ref{DomEq}) explicitly, task which proves to be difficult most of the times. The following change of coordinates is made:
\begin{equation}
\label{VarPhi}
f \longrightarrow \Psi = \Psi(f)
\end{equation}
where $\Psi(f)$ is the domain wall solution of the field theory and thus $f$ is given in terms of the inverse function of $\Psi(f)$. This leads to the following change for integration variables:
\begin{equation}
\label{VarPhi}
\int dx dy dz ...\longrightarrow \int dy dz \int {df \over \lambda(y,z)} ...
\longrightarrow \int {{dy dz} \over \lambda(y,z)} \int {d\Psi \over \Psi'
 }...
\end{equation}
where $\Psi'(f)$ is expressed in terms of $\Psi$ in equations (\ref{deriv1}), (\ref{deriv2}). Using equation (\ref{FreeSS}) and the limiting procedure (\ref{VarPhi}) results in:
\begin{eqnarray}
F[\lambda; \sigma] &=&\int d\Sigma \bigg(A_1 \Big(\lambda^{-1}+\lambda \Big) + A_2 \lambda^3 + \nonumber \\
&& + A_3 \lambda^2 \Big(-H\Big) + A_4 \lambda H^2 + A_5 \lambda \Big(H^2 - {K \over 2}\Big)\bigg) \label{FSS} 
\end{eqnarray}
where $\sigma(y,z)$ is a parameterization of the interface and $\lambda(y,z)$ is the inverse of the interfacial width in the normal direction, and $d\Sigma$ is the geometric measure of integration. The expression for coefficients after the change of variable to $\Psi$, $A_{1..5}$, are:
\begin{eqnarray}
A_1&=& \int_{\Psi_{-}}^{\Psi_{+}} d\Psi M(\Psi) \Psi' \nonumber\\
A_2&=& \int_{\Psi_{-}}^{\Psi_{+}} d\Psi \bigg( N(\Psi) \Psi'^3 + \Big(P(\Psi) + Q(\Psi)\Big)\Psi'\Psi'' + \Big( R(\Psi) + S(\Psi) \Big) {\Psi''^2 \over \Psi'} \bigg)\nonumber \\
A_3&=& 2 \int_{\Psi_{-}}^{\Psi_{+}} d\Psi \Big( P(\Psi) \Psi'^2 + 2 R(\Psi) \Psi'' \Big) \label{AAA} \\
A_4&=& 4 \int_{\Psi_{-}}^{\Psi_{+}} d\Psi  R(\Psi) \Psi' \nonumber\\
A_5&=& 4 \int_{\Psi_{-}}^{\Psi_{+}} d\Psi  S(\Psi) \Psi' \nonumber
\end{eqnarray}
where $M$, $N$, $P$, $Q$, $R$ and $S$  are the functions of $\phi$ in the original free energy (\ref{action}) and the  $\Psi'$ and $\Psi''$ denote the first order and second order derivatives of the $\Psi(f)$ in terms of $\Psi$. All these coefficients are defined with the normal to the surface oriented from the phase having $\Psi_{-}$ to the one with $\Psi_{+}$. The identification of parameters using equation (\ref{HelfrichS}) and (\ref{FSS}) gives:
\begin{eqnarray}
\gamma & = & A_1 \Big(\lambda^{-1}+\lambda \Big) + A_2 \lambda^3 \label{HefSS1} \\
c_H & = & \Big( A_4 + A_5 \Big) \lambda \label{HefSS2}\\
c_G & = & - {A_5 \over 2} \lambda \label{HefSS3}\\
H_0 & = & { A_3 \over {2 ( A_4 + A_5) }} \lambda \label{HefSS4}  
\end{eqnarray}

These expressions connect the free energy coefficients generically present in the equation (\ref{action}) and the interfacial expression proposed phenomenologically in (\ref{HelfrichS}). When the additional derivative terms are absent, the free energy reduces to the first term of $A_1$ leaving the second term of $A_1$ and $A_{2..5}$ zero. Only a contribution from interfacial area and only the interfacial tension parameter is nonzero. In this case, minimizing free energy (\ref{AAA}) with respect to $\lambda$ leads to constant $\lambda = 1$ independent of the details of the interface, making $\lambda$ a redundant field from the point of view of any interfacial dynamics. Furthermore, in the absence of higher derivative terms in the effective free energy expression, it is not possible to achieve variations in interfacial width underlining joining and splitting of interfaces. The interfacial tension obtained after fixing $\lambda=1$ is $\gamma = 2 A_1$. In the case treated in section 5, homopolymers mixture, agreement is obtained with the one proposed in \cite{helfand}. 

The proper way of deriving the equilibrium equations is by varying the metric and the interfacial width (namely, $\sigma(y,z)$) and $\lambda(y,z)$). The resulting equations are not transparent and other types of variations, though not fully rigorous, are employed further. The equations below are obtained by variation of free energy in terms of $\lambda$ alone. The resulting equation describes the equilibrium conformation (minimum of the free energy) when geometric properties of the interface ($H$, $K$) are given and it is used to evaluate the interfacial width: 
\begin{equation}
A_1(-\lambda^{-2}+1)+3 A_2 \lambda^2+ 2 A_3 \lambda (-H) + ((A_4 + A_5) H^2 - {A_5 \over 2} K) = 0  \label{IntEq0} 
\end{equation}
where $A_{2..5}$ are correction terms as compared to $A_1$. The equilibrium interfacial width (\ref{FSS}), depends on the values of curvature invariants. For clarity, the discussion is focused on the interplay between the interfacial width and only one of the curvature invariants, namely mean curvature $H$. This is achieved by an additional condition, $S(\phi) = 0$ in equation (\ref{action}), leading to $A_5 = 0$ in equation (\ref{FSS}). Variation of free energy with respect to both $\lambda$ and $H$ gives:
\begin{eqnarray}
&  A_1(-\lambda^{-2}+1)+3 A_2 \lambda^2+ 2 A_3 \lambda (-H) +  A_4 H^2  = 0  \label{IntEq} \\
& -A_3\lambda^2 + 2 A_4 \lambda H =  0 \label{IntEq1}
\end{eqnarray}
Solving the equation (\ref{IntEq1}) with respect with $H$ and replacing it back in equation (\ref{IntEq}) leads to:
\begin{eqnarray}
&H = {{A_3}\over{2A_4}} \lambda \\
&-\lambda^{-2}+ 1 + 3\left({4 A_2 A_4-{A_3}^2}\over{4 A_1 A_4}\right)\lambda^2  =  0
\end{eqnarray}
The exact solution is given below:
\begin{eqnarray}
a &=& 3\left({4 A_2 A_4-{A_3}^2}\over{4 A_1 A_4}\right)\nonumber\\
\lambda &=& \sqrt{{-1 + \sqrt{1+4a}}\over{2a}}\nonumber\\ 
H &=& {{A_3}\over{2A_4}}\lambda \nonumber
\end{eqnarray}
An approximate zeroth order solution is more relevant for the following discussion. The coefficients accompanying the higher derivative terms are small corrections to the zeroth order term in the free energy. In zeroth approximation, the terms containing $A_{2..4}$ lead to the conclusion that the constant $a$ is a first order correction, giving:
\begin{eqnarray}
\lambda & = & 1\nonumber\\
H & = & {A_3 \over {2 A_4}} \label{HI}
\end{eqnarray} 
Although coefficients $A_3$ and $A_4$ are first order correction terms, their ratio is not necessarily small in first order and that amounts to having a zeroth order nontrivial intrinsic mean curvature of interfaces. Thus, even in the zeroth order approximation, the interface could show a nonzero intrinsic mean curvature. In the first approximation, the inverse of the interfacial width departs from $1$. 
The zeroth order term in the interfacial free energy density constraints $\lambda$ to have values of order $1$ irrespective of the values of the corrective terms in the free energy expression (\ref{action}). In this zeroth order approximation, when $\lambda =1 $, a connection with Helfrich interfacial free energy (\ref{HelfrichS}) can be made using equations (\ref{HefSS1}) to (\ref{HefSS4}).  

The discussion above is valid for the ideal equilibrium configuration, but a realistic dynamics of field theory brings in particular points where curvature values differ significantly from those at equilibrium. Alternatively, the curvature is influenced by boundary conditions, insertion of high curvature boundary surfaces or external chemical potentials. For this reason, the first equation of (\ref{IntEq}) is considered such that the mean curvature $H = H_0$ is fixed to a value different from that found at equilibrium and the interfacial width $\lambda$ is obtained by solving the equation (\ref{IntEq}): 
\begin{equation}
\label{IntCurv}
3 {A_2 \over A_1} \lambda^4 -2 {A_3 \over A_1} H_0 \lambda^3 +(1 + {A_4 \over A_1} H_0^2) \lambda^2 -1 =0
\end{equation}
Two limiting cases are considered: flat surfaces with $H_0 = 0$ and surfaces with high curvature, $H_0 \rightarrow \infty$:   
\begin{eqnarray}
H_0 &\rightarrow& 0 : \lambda \rightarrow \left(A_1 {{\left( 1+12 {A_2 \over A_1} \right)^{1/2}-1} \over {6 A_2}}\right)^{1/2} \\
H_0 &\rightarrow& \pm \infty : \lambda \rightarrow \left({A_1 \over A_4}\right)^{1/2} {1 \over  |H_0|} \label{LH}
\end{eqnarray}
In particular, for the case when $A_2 \ll A_1$, the limit $H_0 \rightarrow 0$ gives $\lambda \rightarrow 1$ and the interfacial tension of the flat interface is $2 A_1$. The limit $H \rightarrow \infty$ shows that the interfacial width, $\lambda^{-1}$  goes to $\infty$, thus at high curvature points, the interface becomes smoother and the phases start mixing along the normal direction to the interface. Thus, high curvature-points are smoothed out by an increase of interfacial width to $\infty$ ($\lambda \rightarrow 0$) leading to the conclusion that higher derivative terms do not allow exposed high-curvature interfaces to appear dynamically.  

\section{A real system - The binary polymeric mixture}
A particular free energy expression that contains higher derivative terms is considered in this section. This expression comes from a mixture of two polymeric components, having the same Kuhn length $a$, and different polymeric lengths $N_A$, $N_B$ interacting with a coupling constant, Flory Huggins parameter, $\chi$. A free energy of type (\ref{action}) is derived in \cite{freed} for the binary incompressible mixture of homopolymers by a formal expansion of free energy in terms of $R_g^2 \nabla^2$, where $R_g^2={N a^2 \over 6}$ represent either polymer radius of gyration (when $N$ is replaced by particular polymeric lengths $N_{A,B}$). The formal expansion in powers of $R_g^2 \nabla^2$ is cut off at order 2, $O(2)$, in the expansion term and the following expression is obtained, see also \cite{lifschitz}:
\begin{eqnarray}
F[\phi] & = & k_B T \int dr \bigg( V_{FH}(\phi)+{{a^2 (\nabla \phi)^2} \over { 36 \phi (1-\phi)}}+ {{a^4 (\nabla \phi)^2 \dd \phi} \over 6480} \Big( {N_B \over (1-\phi)^2}-{N_A \over \phi^2} \Big)+ \nonumber\\
& & \qquad\qquad\qquad\qquad + {{a^4 (\dd \phi)^2} \over 2592} \Big( {N_A \over \phi} + {N_B \over {1-\phi}} \Big) \bigg) \label{polyaction}
\end{eqnarray}
where $k_B$ is the Boltzmann constant and $T$ is the temperature. $V_{FH}$ the Flory-Huggins potential:
\begin{equation}
V_{FH}(\phi)={{\phi \ln(\phi)}\over N_A} + {{(1-\phi) \ln(1-\phi)} \over N_B} + \chi \phi (1-\phi)
\end{equation}
The enthalpic term, $\chi \phi (1-\phi)$ is contained in the Flory-Huggins potential all the other terms being of entropic origin. Entropic terms describe two types of entropies. The translational entropy is described by terms of $V_{FH}$ depending of $N_A$, $N_B$ and the regular $(\nabla \phi)^2$ containing term. The entropy associated with the internal degrees of freedom and in particular with polymer "stretching" ability is represented by higher derivative terms in the free energy expression. For given polymeric lengths, a critical value for the enthalpic parameter $\chi_{critical}$ is calculated:
\begin{equation}
\label{ChiCrtical}
\chi_{critical}={1 \over 2} \left({1 \over \sqrt{N_A}}+{1 \over \sqrt{N_B}}\right)^2
\end{equation} 
For values of $\chi$ smaller than $\chi_{critical}$, the Flory-Huggins potential, $V_{FH}$ has a single minimum and describes a homogeneous polymeric mixture, whereas for values of $\chi$ greater than $\chi_{critical}$, phase segregation is possible and in fact $V_{FH}$ develops two distinct minima. 
Another feature of the above free energy is its dependence on molecular weight of the polymers. As the molecular weight of homopolymers increases, there is a redistribution of the entropic free energy consisting in a decrease in translational entropy term, multiplied by ${1\over N_{A,B}}$ factors, and an increase in internal entropy, multiplied by $N_{A,B}$ factors. The entropic energy redistribution is consistent with the decrease in translational entropy as the molecular weight increases and an increase in internal degrees of freedom entropy as the molecular weight increases. 

The Flory-Huggins potential is asymmetric when $N_A \neq N_B$ and the two minima have an energy difference between them resulting in a bulk contribution to the free energy. The analysis of the bulk contribution is beyond the goal of this analysis and appropriate physical limits will be assumed to reduce the discussion to interface alone. When $\chi$ is large enough (an order of magnitude greater than $\chi_{critical}$ or roughly, $\chi N_A \gg 2$ and $\chi N_B \gg 2$), the strong segregation limit is valid and although $V_{FH}$ is still asymmetric, it presents two approximately equal minima localized at $\Psi_{-} \approx 0$ and $\Psi_{+} \approx 1$. In addition, since expressions from section 4 are used, the assumption of negligible variations in $\lambda$ is valid.  Within such a model, the following expressions are employed to calculate $A$'s in equations (\ref{AAA}):
\begin{eqnarray}
V(\phi)& = & V_{FH}(\phi) \label{VP1}\\
M(\phi)& = & {a^2 \over 36} {1\over {\phi (1-\phi)}}\label{VP2}\\
N(\phi)& = & Q(\phi) = S(\phi)  = 0 \label{VP3}\\ 
P(\phi)& = & {a^4 \over 6480 } \left( - { N_A \over \phi^2}+ { N_B \over (1 - \phi)^2} \right) \label{VP4}\\
R(\phi)& = & {a^4 \over 2592 } \left( { N_A \over \phi} + { N_B \over {1 - \phi}}\right)\label{VP5}
\end{eqnarray}
The strong segregation limit leads to $\Psi_{-,+}$ being $0$ and correspondingly, $1$ and all expressions are Taylor expanded in powers of ${1 \over {\chi N_{A,B}}}$. The formulas developed in section 4 are applied to this realistic free energy and the analytic expressions are obtained using an expansion in $1 \over {\chi N_{A,B}}$. The values of $\Psi_{+}$ and $\Psi_{-}$ close to $0$ and $1$ allow analytic integration over $\Psi$ for the expressions derived in this section. Calculations are done in the first two relevant orders for each $A$. Only $V_{FH}$ depends on $\chi$, and the potential is rewritten in a form more appropriate for perturbation:
\begin{equation}
V_{FH}(\phi)=\chi \phi (1-\phi) \left( 1 +  {1\over {\chi N_A}} {\ln(\phi)\over (1-\phi)}+{1\over {\chi N_B}} {\ln(1-\phi)\over \phi} \right)
\end{equation}
The expression is exact and it is reflected in a perturbation fashion in the following expressions: $\Psi'$, ${1 \over \Psi'}$ and $\Psi''^2$, which appear in (\ref{AAA}). These functions are treated as function of $\Psi$ using (\ref{deriv1}), (\ref{deriv2}) and (\ref{VP1}) to (\ref{VP5}): 
\begin{eqnarray}
\Psi'& \approx &{{6 \sqrt{\chi}} \over a} \Psi (1-\Psi) \bigg(1 + {1 \over {2 \chi N_B}} {\ln(1-\Psi) \over \Psi} + {1 \over {2 \chi N_A}} {\ln(\Psi) \over {1-\Psi}}\bigg) \nonumber\\
{1 \over \Psi'}& \approx & { a \over {6 \sqrt{\chi}} } {1 \over {\Psi (1-\Psi)}} \bigg(1 - {1 \over {2 \chi N_B}} {\ln(1-\Psi) \over \Psi} - {1 \over {2 \chi N_A}} {\ln(\Psi) \over {1-\Psi}}\bigg) \nonumber \\
\Psi'' & \approx & {{18 \chi} \over a^2} \Psi (1-\Psi) \bigg( (2-4\Psi) +  {1 \over {\chi N_A}} - {1 \over {\chi N_B}} + {{(2-3\Psi) \over {\chi N_A}}}{{\ln(\Psi)}\over {1-\Psi}}+ \nonumber \\ 
&& \qquad\qquad + {(1-3\Psi) \over {\chi N_B}}{{\ln(1-\Psi)}\over \Psi}\bigg) \nonumber \\
\Psi''^2 &\approx& {{18^2 \chi^2} \over a^4} \Psi^2 (1-\Psi)^2 \bigg( (2-4\Psi)^2 + 2 (2-4\Psi) \times \nonumber \\
& &  \qquad \qquad \times \Big({1 \over {\chi N_A}} -{ 1 \over {\chi N_B}} + {{(2-3\Psi) \over {\chi N_A}}}{{\ln(\Psi)}\over {1-\Psi}}+{{(1-3\Psi) \over {\chi N_B}}}{{\ln(1-\Psi)}\over \Psi}\Big)\bigg) \nonumber
\end{eqnarray} 
The corresponding coefficients in their first two orders in ${1 \over {\chi N_{A,B}}}$ are given below after evaluation of integrals (\ref{AAA}) using Mathematica \cite{mathematica} and the equations above: 
\begin{eqnarray}
A_1&=&{1 \over 6} a \chi^{1/2} \left(1-{\pi^2 \over 12}{1\over \chi}\left({1 \over N_A}+{1 \over N_B}\right)\right) \nonumber \\
A_2&=&{1\over 120} a \chi^{1/2} \left( \chi (N_A + N_B) -{17 \over 3}-{{1+\pi^2}\over 6}({N_A \over N_B}+{N_B \over N_A}) \right) \nonumber \\
A_3&=&{1 \over 180} a^2 (\chi (N_A-N_B) + ({15 \over 4}-{\pi^2 \over 2})({N_A \over N_B}-{N_B \over N_A})) \nonumber \\
A_4&=&{1 \over 864} a^3 \chi^{-1/2} \left(\chi(N_A+N_B)-2-(\pi^2/6-1)({N_A \over N_B} + {N_B \over N_A})\right) \nonumber
\end{eqnarray}
The first term OF $A_1$ was derived $\cite{helfrich}$ for a free energy having only a quadratic derivative term with a different numeric coefficient and without higher derivative terms. For expressions above only the first two order terms of each $A$ are considered. For example the intrinsic mean curvature (\ref{HI}) is 
\begin{equation}
H = {12 \over 5} {1\over {a \sqrt{\chi}}} {{N_A - N_B}\over{N_A + N_B}}
\end{equation}
The most important parameter controlling the intrinsic curvature is the ratio between homopolymers molecular weights: ${N_A \over N_B}$. The closer this number is to $0$, namely the larger the difference between the molecular weights the highest the value, the higher the intrinsic curvature of the interface. Also, the interface curves towards the higher molecular weight polymer. Similarly, there is no intrinsic curvature for a system where $N_A=N_B$. Another conclusion is that lower values of Kuhn's parameter, $a$ and enthalpic parameter, $\chi$ promote higher values of intrinsic curvature. In the approximation used in this work, the low values achieved by $\chi$ are quite limited since $\chi > \chi_{critical}$, whereas low values of $a$ are more accessible. As $a$ decreases, the steepness of the interfacial profile $\Psi(x)$ increases. Such increase, at some point, would contradict the formal perturbation expansion in $R_g^2 \nabla^2$. Lowering $a$ will result in steeper interfaces increasing the weight of higher derivative terms and finally shifting the contributions from translational entropy to internal entropy. Another conclusion regards the dependence of interfacial width of curvature. Dependence of interfacial width on mean curvature for large values of curvature $H_0$ is (\ref{LH}):
\begin{equation}
\lambda \rightarrow {144 \over a^2} {1 \over {\chi (N_A + N_B)}} { 1 \over |H_0|}
\end{equation}
The expression above shows that sensitivity of the interfacial width to interfacial curvature is increased for low values of Kuhn's parameter, $a$, and also for low values of $ \chi (N_A + N_B) $ (within the lower limit $\chi_{critical}$).      
The results highlight the intermediate regime of interest. Reaching the exact parameter region of interest in general form is difficult, but detailed parameter ranges can be obtained for particular cases. The difficulty for reaching an exact parameter range stems from the regime that is studied: the intermediate range where both translation and extension entropy play a role. The following general conclusions can be reached. Lower values of $\chi$, closer to $\chi_{critical}$ as well as lower values of Kuhn's length will be more favorable for interfacial modifications. By contrast, high values for $\chi$ and for Kuhn's length, $a$, decrease sensitivity increasing stability of interfaces. Larger discrepancy in molecular weights $N_A$ and $N_B$ will promote interfaces with higher curvature oriented towards the phase with higher molecular weight.

\section{Conclusions}
In this work, high derivative free energy terms (higher than $\nabla^2$) for phase separated systems (systems having domain-wall solutions) are analyzed. These terms are directly associated with geometric invariants, such as mean and Gaussian curvature terms in interfacial free energy. It is shown that these higher derivative terms allow an energetic interaction of geometry of the interface with the interfacial profile. In particular, it is shown that interfacial curvatures affect interfacial width, giving thus a mechanism that can be used to achieve in a continuous manner the joining and splitting of interfaces. The method is applied to a realistic free energy used in homopolymers blends and qualitative results about the parameter range most appropriate for interfacial modifications are presented. In the case of homopolymer blends, higher derivative terms come from the entropy of "stretching" and it is shown that they affect directly the geometry of the interface.
 
Finally, this work shows the potential benefits of introducing higher derivative terms in field free energy models used in computational modeling of chemical systems exhibiting phase separation and rich membrane-like structures \cite{balazs}. The terms are potentially useful in tackling other types of systems where interfaces and curvature contributions appear at molecular scales.

\section{Appendix 1 - Geometric free energy terms}
In this section the expressions used in section 2 are derived and the connection between fields living on the three dimensional space and those living on two dimensional interfacial space is established. Following the procedure described in section 1, the three dimensional field theory is written, using a field redefinition, in terms of field $f$. The field $f$ has a direct relation with the interface (the surface where $f=0$). Considering a point on the interface, a coordinate system can be chosen such as the interfacial point under study is $(0,0,0)$ and a physical Ansatz for the three dimensional field $f$ is given in section 2. The function is given by $f(x,y,z)=\lambda(y,z) (x - \sigma(y,z))$, where $\lambda$ describes the inverse of width of the interface at the interfacial point $(y,z)=(0,0)$ and $\sigma$ describes the location of the interface.  Following this Ansatz, the equation for the interface in the neighborhood of interface is $f(x,y,z)=0$ or equivalently, $x=\sigma(y,z)$. Since $(0,0,0)$ is an interfacial point, $\sigma(0,0)=0$ and by an orientation choice of the coordinate system, $\partial_y \sigma = \partial_z \sigma = 0$. The $y$ and $z$ coordinates are tangent to the interface at point $(0,0,0)$ and the $x$ axis is normal to the interface. Using the equation for the interface, the expressions of the first fundamental form (metric, $ds^2$) and the second fundamental form ($\Pi_{\alpha \beta}$) of the interface at point $(0,0,0)$ are computed using Monge representation in terms of the field $\sigma$ (for additional details see \cite{Safran}): 
\begin{eqnarray}
ds^2&=&\Big( 1+(\partial_y \sigma)^2 \Big) dy^2 + 2 \partial_y \sigma \partial_z \sigma dy dz+ \Big( 1+(\partial_z \sigma)^2 \Big) dz^2 \nonumber \\
&&\qquad \qquad \qquad \qquad \stackrel{(y,z) \rightarrow 0} {\longrightarrow}dy^2+ dz^2 \nonumber\\
\Pi_{yy}&=& \frac{\partial_{yy}\sigma}{\sqrt{1+(\partial_y \sigma)^2+(\partial_z \sigma)^2}}
\stackrel{(y,z) \rightarrow 0}{\longrightarrow}\partial_{yy}\sigma|_0 \nonumber\\
\Pi_{zz}&=&\frac{\partial_{zz}\sigma}{\sqrt{1+(\partial_y \sigma)^2+(\partial_z \sigma)^2}}
\stackrel{(y,z) \rightarrow 0}{\longrightarrow}\partial_{zz}\sigma|_0\nonumber\\
\Pi_{yz}&=&\frac{\partial_{yz}\sigma}{\sqrt{1+(\partial_y \sigma)^2+(\partial_z \sigma)^2}}
\stackrel{(y,z) \rightarrow 0}{\longrightarrow}\partial_{yz}\sigma|_0 \nonumber
\end{eqnarray}
In the same limit, $(y,z) \rightarrow 0$ on the interface, the expressions for the 
curvature scalar invariants, mean curvature $H = {1 \over 2} Tr (\Pi)$ and 
Gaussian curvature $K = Det(\Pi)$, are given below:
\begin{eqnarray}
H & \stackrel{(y,z) \rightarrow 0}{\longrightarrow}& {1 \over 2}  \; {\dd \sigma|_0}\\
K & \stackrel{(y,z) \rightarrow 0}{\longrightarrow}& (\partial_{yy}\sigma  \; \partial_{zz}\sigma-(\partial_{yz}\sigma)^2)|_0 
\end{eqnarray}
In order to make contact with the terms appearing in the three dimensional field theory, 
all quantities in the three dimensional field theory are computed in the limit 
$(y,z) \rightarrow 0$ with the normal coordinate, $x$, being arbitrary (still in the neighborhood of the interfacial point $(0,0,0)$). The derivatives of the field $f$ in this limit are shown below:
\begin{eqnarray}
\partial_x f(x,y,z)&=&\lambda \stackrel{(y,z)\rightarrow 0}{\longrightarrow} \lambda |_0 \nonumber\\
\partial_y f(x,y,z)&=& \partial_y \lambda \; (x-\sigma)-\lambda \; \partial_y \sigma 
\stackrel{(y,z)\rightarrow 0}{\longrightarrow} x \; \partial_y \lambda |_0 \nonumber\\    
\partial_z f(x,y,z)&=& \partial_z \lambda\; (x-\sigma)-\lambda  \; \partial_z \sigma 
\stackrel{(y,z)\rightarrow 0}{\longrightarrow} x  \; \partial_z \lambda |_0\nonumber\\    
\partial_{xx} f(x,y,z)&=& 0\nonumber\\
\partial_{yy} f(x,y,z)&=& \partial_{yy} \lambda  \; (x -\sigma) - 2  \; \partial_y \lambda  \; \partial_y \sigma)-\lambda  \; \partial_{yy}\sigma 
\stackrel{(y,z)\rightarrow 0}{\longrightarrow} (x  \; \partial_{yy} \lambda - \lambda  \; \Pi_{yy})|_0 \nonumber \\
\partial_{zz} f(x,y,z)&=& \partial_{zz} \lambda  \; (x - \sigma) - 2  \; \partial_z \lambda  \; \partial_z \sigma-\lambda  \; \partial_{zz}\sigma
\stackrel{(y,z)\rightarrow 0}{\longrightarrow} (x  \; \partial_{zz} \lambda - \lambda  \; \Pi_{zz})|_0 \nonumber\\
\partial_{xy} f(x,y,z)&=& \partial_y \lambda \stackrel{(y,z)\rightarrow 0}{\longrightarrow} \partial_y \lambda |_0 \nonumber\\
\partial_{xz} f(x,y,z)&=& \partial_z \lambda \stackrel{(y,z)\rightarrow 0}{\longrightarrow} \partial_z \lambda |_0\nonumber\\
\partial_{yz} f(x,y,z)&=& \partial_{yz} \lambda  \; (x - \sigma) - \partial_y \lambda  \; \partial_z \sigma -
\partial_z \lambda \partial_y \sigma - \lambda \partial_{yz} \sigma \nonumber\\
&& \qquad \qquad \qquad \stackrel{(y,z)\rightarrow 0}{\longrightarrow} (x  \; \partial_{yz} \lambda - \lambda  \; \Pi_{yz})|_0 \nonumber
\end{eqnarray}
where $\Pi_{\alpha\beta}$ is the second fundamental tensor of the interfacial surface. In particular the Laplacian of the field $f$ is connected with the mean curvature scalar $H$.  
\begin{equation}
\dd f(x,y,z)\stackrel{(y,z)\rightarrow 0}{\longrightarrow} (x  \; \dd \lambda - 2  \; \lambda  \; H ) |_0
\end{equation}
Using the equations above, each of the terms in the free energy density can be connected to interfacial fields such as interfacial depth $\lambda$ and interfacial metric (namely, $\sigma$):
\begin{eqnarray}
(\nabla f)^2 & \stackrel{(y,z)\rightarrow 0}{\longrightarrow} & (\lambda^2 + x^2 (\nabla \lambda)^2 )|_0 \nonumber \\ 
(\nabla f)^4 & \stackrel{(y,z)\rightarrow 0}{\longrightarrow} & (\lambda^2 + x^2 (\nabla \lambda)^2 )^2|_0 \nonumber \\
(\nabla f)^2 \dd f & \stackrel{(y,z)\rightarrow 0}{\longrightarrow}&
(\lambda^2 + x^2  (\nabla \lambda)^2 )(x   \dd \lambda - 2  \lambda   H) |_0 \nonumber\\
\partial_i f \partial_j f \partial^{ij} f & \stackrel{(y,z)\rightarrow 0}{\longrightarrow} &
x  \lambda  (\nabla \lambda)^2 + x^3  \partial_{\alpha}  \lambda \partial_{\beta} \lambda  \partial^{\alpha\beta}\lambda - x^2 \lambda \partial_{\alpha} \lambda \partial_{\beta} \lambda  \Pi^{\alpha\beta} |_0 \nonumber\\
(\dd f)^2 & \stackrel{(y,z)\rightarrow 0}{\longrightarrow} & (x \dd \lambda - 2 \lambda H)^2 |_0 \nonumber\\ 
\partial_{ij} f \partial^{ij} f & \stackrel{(y,z)\rightarrow 0}{\longrightarrow} &
x^2 \partial_{\alpha \beta} \lambda \partial^{\alpha \beta} \lambda + 2 (\nabla \lambda)^2 -  2 x \lambda \partial _{\alpha \beta} \lambda \Pi^{\alpha \beta} + \lambda^2 \Pi_{\alpha \beta} \Pi^{\alpha \beta}|_0 \nonumber
\end{eqnarray}
where the derivatives, in terms of $\nabla$, are understood to act the space where each function is defined (i.e. if the field is $f$, then three dimensional $\nabla$ and $\dd$
are used, whereas if the field is $\lambda$ or $\sigma$ two dimensional $\nabla$ and $\dd$ 
are used). In addition for the final expressions used in section 2, $x$ on the right side is replaced by ${f \over \lambda}$. The metric on the three dimensional space is flat and thus no distinction appears between upper and lower indices. The final general expression shows a geometric invariant character. For this reason, the fully geometric invariant expression of each term was highlighted on the right side of the equations above. In addition, the term $\Pi_{\alpha \beta}  \; \Pi^{\alpha \beta}$ can be expressed in terms of the curvature invariants $H$ and $K$ as shown in the equation (\ref{HK1}), (\ref{HK2}) of section 2.

\section{Appendix 2 - Dimensional reduction}  
In this section, the formulas used in section 3 are derived. The free energy action in three dimensions is reduced in the normal direction on the interface. In order to simplify the action given in the equation (\ref{P4action}), within the general setup given in (\ref{FreeS}), a few properties for the domain-wall solution of $\phi^4$ potential of section 3 are added :
\begin{eqnarray}
\Psi(x) & = & \tanh(x) \nonumber \\
\Psi'(x) & = & 1 - \tanh(x)^2 \\
\Psi''(x) & = & -2 \tanh(x) \left( 1 - \tanh(x)^2 \right) \nonumber 
\end{eqnarray}
For the first term in (\ref{P4action}), the term that depends on the coefficient $M(\Psi)={1 \over 2}$ in equation (\ref{FreeS}), two integrals are evaluated and the results obtained using Mathematica \cite{mathematica} are listed below:
\begin{eqnarray}
\int_{-\infty}^{+\infty} df \Psi'^2  = {4 \over 3};  \;\;  \int_{-\infty}^{+\infty} df f \Psi' = 0  
\;\; \int_{-\infty}^{+\infty} df f^2 \Psi'^2 = {{\pi^2 - 6} \over 9} 
\nonumber
\end{eqnarray}
leading to equation (\ref{NoCorr}). Similarly, for the second term in (\ref{P4action}), the term that depends on the coefficient $N$, three integrals are evaluated and the results are listed below:
\begin{eqnarray}
\int_{-\infty}^{+\infty} df \Psi'^4  &=& {32 \over 35};  \;\;  \int_{-\infty}^{+\infty} df f^2 \Psi'^4 = {{4 (6 \pi^2 - 49)} \over 315} \nonumber \\
\int_{-\infty}^{+\infty} df f^4 \Psi'^4 &=& {{2 (3 \pi^4 - 35 \pi^2 + 60)} \over 225} \nonumber
\end{eqnarray}
with the result given in (\ref{NT}). The third term in (\ref{P4action}), the term that depends on the coefficient $P$, is written in (\ref{PT}) after the evaluation of seven integrals:
\begin{eqnarray}
\int_{-\infty}^{+\infty} df \Psi'^2 \Psi'' &=& \int_{-\infty}^{+\infty} df f^2 \Psi'^2 \Psi'' f^2 = \int_{-\infty}^{+\infty} df f^4 \Psi'^2 \Psi'' = 0 \nonumber\\
\int_{-\infty}^{+\infty} df f \Psi'^3 &=& \int_{-\infty}^{+\infty} df f^3 \Psi'^3 =0  \nonumber \\
\int_{-\infty}^{+\infty} df \Psi'^3 &=& { 16 \over 15}; \;\;  \int_{-\infty}^{+\infty} df f^2 \Psi'^3 = { {4 \pi^2 - 30}  \over 15} \nonumber
\end{eqnarray}
Finally, the last term from (\ref{P4action}), the term that depends on the coefficient $R$, comes with a contribution to the interfacial free energy given by (\ref{RT}) after the removal of a total derivative term and evaluation of additional six integrations: 
\begin{eqnarray}
\int_{-\infty}^{+\infty} df \Psi''^2 &=& { 16 \over 15}, \;\; \int_{-\infty}^{+\infty} df f^2 \Psi''^2 = { {4 \pi^2} \over 45}, \;\; \int_{-\infty}^{+\infty} df f^4 \Psi''^2 = { { 7 \pi^4 - 360 } \over 225} \nonumber \\
\int_{-\infty}^{+\infty} df  \Psi' \Psi'' &=& \int_{-\infty}^{+\infty} df f^2 \Psi' \Psi'' = 0; \;\; \int_{-\infty}^{+\infty} df f \Psi' \Psi'' = - { 2 \over 3} \nonumber \\
\int_{-\infty}^{+\infty} df f^3 \Psi' \Psi'' &=& { {6 - \pi^2} \over 6} \nonumber
\end{eqnarray}

\end{document}